\documentclass[sigconf, natbib=false]{acmart}
\AtBeginDocument{%
  }

\setcopyright{acmlicensed}
\copyrightyear{2018}
\acmYear{2018}
\acmDOI{XXXXXXX.XXXXXXX}
\acmConference[Conference acronym 'XX]{Make sure to enter the correct
  conference title from your rights confirmation email}{June 03--05,
  2018}{Woodstock, NY}
\acmISBN{978-1-4503-XXXX-X/2018/06}

\RequirePackage[
  datamodel=acmdatamodel,
  style=acmnumeric,
  ]{biblatex}

\usepackage{makecell}
\usepackage{booktabs}
\usepackage{subcaption}
\usepackage{caption}

\begin{document}

\title{A Keyframe-Based Approach for Auditing Bias in YouTube Shorts Recommendations}

\author{Mert Can Cakmak}
\affiliation{%
  \institution{Computer and Information Science \\ University of Arkansas - Little Rock}
  \city{Little Rock}
  \state{Arkansas}
  \country{USA}
}
\email{mccakmak@ualr.edu}

\author{Nitin Agarwal}
\affiliation{%
  \institution{ICSI, University of California, Berkeley}
  \city{Berkeley}
  \state{California}
  \country{USA}
}
\affiliation{%
  \institution{COSMOS Research Center \\ University of Arkansas - Little Rock}
  \city{Little Rock}
  \state{Arkansas}
  \country{USA}
}
\email{nxagarwal@ualr.edu}


\renewcommand{\shortauthors}{Cakmak \& Agarwal}

\begin{abstract}
YouTube Shorts and other short-form video platforms now influence how billions engage with content, yet their recommendation systems remain largely opaque. Small shifts in promoted content can significantly impact user exposure, especially for politically sensitive topics. In this work, we propose a keyframe-based method to audit bias and drift in short-form video recommendations. Rather than analyzing full videos or relying on metadata, we extract perceptually salient keyframes, generate captions, and embed both into a shared content space. Using visual mapping across recommendation chains, we observe consistent shifts and clustering patterns that indicate topic drift and potential filtering. Comparing politically sensitive topics with general YouTube categories, we find notable differences in recommendation behavior. Our findings show that keyframes provide an efficient and interpretable lens for understanding bias in short-form video algorithms.

\end{abstract}



\received{20 February 2007}
\received[revised]{12 March 2009}
\received[accepted]{5 June 2009}

\maketitle

\section{Introduction}

Short-form video platforms like YouTube Shorts have rapidly reshaped online content consumption, offering users a continuous, swipe-driven stream of personalized video recommendations. Since its global launch in 2021, YouTube Shorts has grown to over two billion monthly users, making it one of the most influential content delivery systems worldwide. As algorithms decide what users see next, even small recommendation biases can significantly affect exposure to underrepresented or sensitive topics.

Prior studies have highlighted ideological skew, homogenization, and popularity bias in recommender systems, often through metadata or user interaction data. However, these approaches overlook the actual visual content users consume, and analyzing entire videos at scale is computationally intensive. This challenge is especially pronounced in short-form ecosystems where content changes rapidly.

To address this, we introduce a keyframe-based method for analyzing recommendation behavior in YouTube Shorts. Using the PRISM framework, we extract perceptually important keyframes from both seed and recommended videos. We then use CLIP to embed these frames, along with captions generated using the Llama-3.2-11B-Vision-Instruct model, and project them into a shared space using UMAP. This allows us to observe how recommendations visually and thematically shift from the original content.

We collect data across two domains: videos related to the politically sensitive “South China Sea” (SCS) and a control group sampled from general YouTube video categories. The SCS topic is frequently subject to geopolitical framing and content moderation, making it a meaningful case for studying potential recommendation bias. In contrast, general content provides a baseline for understanding typical recommendation behavior. This contrast allows us to assess whether algorithmic dynamics differ when political sensitivity is involved.

Our study is guided by the following research questions:

\begin{itemize}
    \item \textbf{RQ1:} Do YouTube Shorts recommendations visually or thematically shift away from their seed content?

    \item \textbf{RQ2:} Can keyframes and their captions reveal patterns in recommended content that suggest potential bias or filtering?

    \item \textbf{RQ3:} Does the degree of content shift differ between politically sensitive topics and general categories?

\end{itemize}

Our findings show that keyframes provide a focused and interpretable lens for studying short-form recommendations. We observe measurable divergence and clustering, with more pronounced shifts in sensitive-topic cases. This work contributes a scalable, content-driven method for auditing algorithmic behavior in modern video platforms.

\section{Literature Review}

YouTube’s recommendation system, powered by deep neural networks, optimizes for predicted watch time but has raised concerns about various biases and drifts \cite{covington2016deep}. These include selection, popularity, and position bias, as well as evolving issues like recommendation drift, algorithmic drift, and content drift \cite{groh2012social,chen2023bias,garapati2025recommender}. Empirical audits reveal ideological and emotional skews, right-leaning accounts receive more radical content, while emotionally negative or morally charged videos are filtered out deeper in the chain \cite{haroon2022youtube,ledwich2019algorithmic,habib2025youtube}. For sensitive topics such as China–Uyghur, recommendations increasingly favor positive sentiment content, suppressing moral critique \cite{cakmak2024bias,cakmak2023investigating}. Network analyses show that recommendation structures reinforce thematic silos, limiting content diversity \cite{kirdemir2021exploring,kirdemir2021examining,kirdemir2021assessing}.

Short-form platforms like YouTube Shorts and TikTok intensify these issues. Studies find rapid content drift from controversial seeds to apolitical or visually neutral recommendations \cite{cakmak2025unpacking}. TikTok audits reveal partisan bias favoring Republican content \cite{ibrahim2025tiktok}, while Kuaishou exhibits duration bias favoring longer videos \cite{zheng2022dvr,zhan2022deconfounding,lin2023tree}. Algorithmic feedback loops further reinforce mainstream narratives, reducing minority exposure \cite{mansoury2020feedback}. Mitigation strategies include re-ranking, causal modeling, and adversarial learning to address ideological, temporal, and duration-based biases in recommendation pipelines \cite{haroon2022youtube,zhan2022deconfounding,zheng2022dvr}. However, most existing efforts rely on full video analysis or engagement traces, with limited focus on visual-semantic dynamics in short-form formats like YouTube Shorts. 

We present the first keyframe-based approach of bias and drift in YouTube Shorts. Unlike prior work that processes full videos or relies solely on metadata, we extract perceptually salient keyframes as concise visual summaries. These serve as effective proxies for comparing content across recommendation chains, revealing shifts in visual framing and content representation. This approach enables a faster and more focused analysis of recommendation behavior in short-form video systems.

\section{Data Collection}

To examine potential algorithmic and recommendation biases in YouTube Shorts, we collected data across two domains: (1) the South China Sea territorial disputes, and (2) general content categories reflecting YouTube’s broader ecosystem. This dual focus enables comparison between politically sensitive and neutral domains, allowing us to assess whether the platform’s algorithm exhibits differential behavior in response to geopolitical versus everyday content.
\subsection{Keyword Selection}

The South China Sea (SCS) dataset was curated through expert consultations with researchers from the Atlantic Council’s Digital Forensic Research Lab, the National University of Singapore, and De La Salle University. Keywords focus on regional disputes, military activity, and environmental issues \cite{atlanticcouncil_dfrlab, nus_oceanlaw_southchinasea}. For the General Content dataset, we followed YouTube’s content taxonomy \cite{entreresource_youtube_categories_2023}, covering a wide spectrum from entertainment and sports to science and education. Table~\ref{tab:keywords} summarizes the keyword sets.

\begin{table}[h]
\small
\caption{Keywords used for data collection. Representative keywords are shown for the South China Sea; full categories are listed for General Content.}
\label{tab:keywords}
\begin{tabular}{@{}>{\raggedright\arraybackslash}p{0.23\columnwidth} >{\raggedright\arraybackslash}p{0.73\columnwidth}@{}}
\toprule
\textbf{Dataset} & \textbf{Keywords} \\
\midrule
South China Sea & South China Sea dispute, China nine dash line, China artificial islands, US Navy South China Sea, China maritime claims, South China Sea latest news, China vs Philippines, South China Sea oil and gas, ASEAN South China Sea talks, China territorial dispute. \\
General Content & Film \& Animation, Autos \& Vehicles, Music, Pets \& Animals, Sports, Travel \& Events, Gaming, People \& Blogs, Comedy, Entertainment, News \& Politics, How-to \& Style, Education, Science \& Technology, Nonprofits \& Activism. \\
\bottomrule
\end{tabular}
\end{table}

\subsection{YouTube Shorts Collection}

Since the YouTube Data API v3 does not support Shorts, we used APIFY’s YouTube Scraper \cite{streamers_2024_youtube} to collect Shorts video IDs based on the keywords in Table~\ref{tab:keywords}. The scraper was configured to retrieve only Shorts by enabling the \texttt{maxResultsShorts} parameter and disabling other video types. We applied no date filters and used the \texttt{relevancy} sort option to prioritize videos most relevant to the keywords. We collected 500 seed videos for each domain: South China Sea and General Content.
\subsection{Recommendation Collection}

To collect Shorts recommendations, we developed a custom scraping framework, as existing tools do not support this functionality. Each seed video was opened in a fresh Selenium-driven browser session with no login, cookies, or browsing history to simulate a neutral user environment. We simulated user interaction by scrolling through recommended Shorts up to a depth of 10, where depth refers to the position in the recommendation chain. The browser was fully reset after each session to avoid cross-session contamination. From each dataset, we collected 5,000 recommended videos. Combined with the 500 seed videos, each dataset contains 5,500 Shorts, totaling 11,000 videos overall.

\subsection{Keyframe Collection}

\begin{figure}[ht]
\centering
\includegraphics[width=0.75\linewidth]{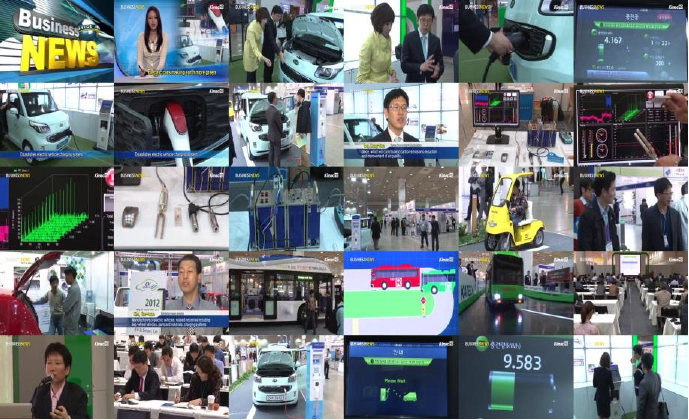}
\caption{Keyframes extracted using the PRISM framework.}
\label{keyframes}
\Description{}
\end{figure}

To extract representative visuals from each YouTube Short, we employed the PRISM framework \cite{cakmak2025prism}, which identifies keyframes based on perceptual changes aligned with human visual sensitivity. The framework takes a video as input and returns visually salient frames that represent important moments. This perceptually guided approach is well-suited for Shorts content, where visual impact and rapid engagement are central to viewer experience. By focusing on human-like perception, PRISM helps surface frames that are more likely to capture user attention and potentially influence recommendation dynamics, especially in politically sensitive or emotionally charged videos. An example of extracted keyframes is shown in Figure~\ref{keyframes}.

We selected PRISM for its strong performance across accuracy, fidelity, and compression efficiency, as summarized in Table~\ref{tab:prism_comparison}, making it a practical and scalable choice for high-volume Shorts analysis.

\begin{table}[ht]
\centering
\caption{Performance comparison of keyframe extraction models with reported accuracy, fidelity, and compression ratio (CR).}
\label{tab:prism_comparison}
\begin{tabular}{lccc}
\toprule
\textbf{Model} & \textbf{Accuracy (\%)} & \textbf{Fidelity (\%)} & \textbf{CR (\%)} \\
\midrule
\textbf{PRISM \cite{cakmak2025prism}}                  & 85.58 & 70.30 & 99.23 \\
LiveLight \cite{zhao2014quasi} & 72.30          & 80.00 & 90.00          \\
DSVS \cite{cong2011towards}    & 66.00          & 75.00 & 95.00          \\
\bottomrule
\end{tabular}
\end{table}

The final number of videos and corresponding keyframes used in this study are summarized in Table~\ref{tab:dataset_stats}. These keyframes form the basis for downstream analysis of algorithmic behavior and content characteristics.

\begin{table}[ht]
\small
\caption{Summary of videos, frames, and extracted keyframes per dataset.}
\label{tab:dataset_stats}
\centering
\begin{tabular}{@{}lccc@{}}
\toprule
\textbf{Dataset} & \textbf{Total Videos} & \textbf{Frames} & \textbf{Keyframes} \\
\midrule
South China Sea & 5,500 & 338,246 & 15,361 \\
General Content & 5,500 & 326,184 & 14,592 \\
\midrule
\textbf{Total} & 11,000 & 664,430 & 29,953 \\
\bottomrule
\end{tabular}
\end{table}

\section{Methodology}
In this section, we describe our approach for generating captions from keyframes and embedding both modalities to analyze YouTube Shorts recommendations.

\subsection{Keyframe Caption Generation}

To support multimodal analysis, we generate natural language captions for each keyframe, providing a complementary semantic signal to the visual content. This aids interpretability and clustering, especially given the dense and fast-paced nature of Shorts.

We use the \texttt{Llama-3.2-11B-Vision-Instruct} \cite{Meta2025Llama3} model, a multimodal, instruction-tuned system designed for tasks like image captioning. It performs competitively in zero-shot settings on benchmarks such as VQAv2 and TextVQA (Table~\ref{tab:model_comparison_vqa}), making it well-suited for producing high-quality captions for short-form content.

\begin{table}[ht]
\centering
\caption{Validation performance of multimodal models on VQAv2 and TextVQA. VQAv2 tests general visual QA, while TextVQA focuses on understanding text within images.}

\begin{tabular}{lcc}
\toprule
\textbf{Model} & \textbf{VQAv2~\cite{goyal2017making}} & \textbf{TextVQA~\cite{singh2019towards}} \\
\midrule
\textbf{LLaMA 3.2 11B~\cite{Meta2025Llama3}}   & 66.8 & 73.1 \\
Flamingo-80B~\cite{alayrac2022flamingo} & 56.3 & 35.0 \\
IDEFICS-9B~\cite{laurencon2023idefics}    & 50.9 & 25.9 \\
\bottomrule
\end{tabular}
\label{tab:model_comparison_vqa}
\end{table}

\subsection{Multimodal Embedding with CLIP}

To convert both keyframes and their generated captions into a unified semantic representation, we used the OpenCLIP implementation of the CLIP ViT-G/14 model \cite{Wortsman2023OpenCLIP}. This model was selected for its state-of-the-art zero-shot performance on vision-language benchmarks (Table~\ref{tab:embedding-pub-benchmarks-full}). Keyframe images were embedded using CLIP’s vision encoder, while the generated captions were embedded using its text encoder. All embeddings were L2-normalized to enable alignment in a shared semantic space. This facilitates downstream tasks such as clustering, where both visual and textual signals contribute to identifying algorithmic patterns. Unlike supervised models trained on fixed label sets, CLIP’s zero-shot approach is more generalizable to open-ended and dynamic content domains like YouTube Shorts, where new trends and topics emerge rapidly.

\begin{table}[ht]
\centering                   
\setlength\tabcolsep{3pt}
\caption{Zero-shot benchmarks for vision-language models.  
Metrics: ImageNet Top-1 (ZS), COCO R@5 (ZS), Flickr30k R@5 (ZS).}
\label{tab:embedding-pub-benchmarks-full}
\resizebox{\columnwidth}{!}{%
  \begin{tabular}{lcccc}
    \toprule
    \textbf{Model} & \#Params 
      & \shortstack{ImageNet~\cite{deng2009imagenet}\\Top-1 (ZS,\%)} 
      & \shortstack{COCO~\cite{lin2014microsoft}\\R@5 (ZS,\%)} 
      & \shortstack{Flickr30k~\cite{young2014image}\\R@5 (ZS,\%)} \\
    \midrule
    \textbf{CLIP ViT-G/14~\cite{radford2021learning,Wortsman2023OpenCLIP}} 
      & 1.8B  & 80.2 & 75.0 & 78.5 \\
    OpenCLIP ViT-H/14~\cite{cherti2023reproducible}      
      & 1.2B  & 78.0 & 73.4 & 75.8 \\
    OpenCLIP RN50×64~\cite{cherti2023reproducible}       
      & 435M  & 70.4 & 62.5 & 63.9 \\
    CLIP ViT-L/14~\cite{radford2021learning}              
      & 427M  & 75.3 & 70.2 & 72.5 \\
    CLIP ViT-B/32~\cite{radford2021learning}              
      & 150M  & 63.3 & 60.5 & 62.0 \\
    ALIGN~\cite{jia2021scaling} 
      & 1.8B  & 76.4 & 77.0 & 79.2 \\
    BLIP~\cite{li2022blip} 
      & 213M  & 74.3 & 72.4 & 74.7 \\
    SigLIP ViT-B/16~\cite{zhou2024scaling} 
      & 86M   & 74.0 & 68.5 & 70.0 \\
    \bottomrule
    \multicolumn{5}{l}{ All metrics are reported in zero-shot settings.}
  \end{tabular}%
}
\end{table}

\section{Results}

We visualize the structure of keyframe and caption embeddings using UMAP \cite{mcinnes2018umap}, a non-linear dimensionality reduction method that preserves meaningful relationships in high-dimensional data. This enables intuitive comparisons between seed and recommended content across domains and modalities.

\begin{figure}[ht]
    \centering
    \begin{subfigure}[b]{0.48\columnwidth}
        \centering
        \includegraphics[width=\linewidth]{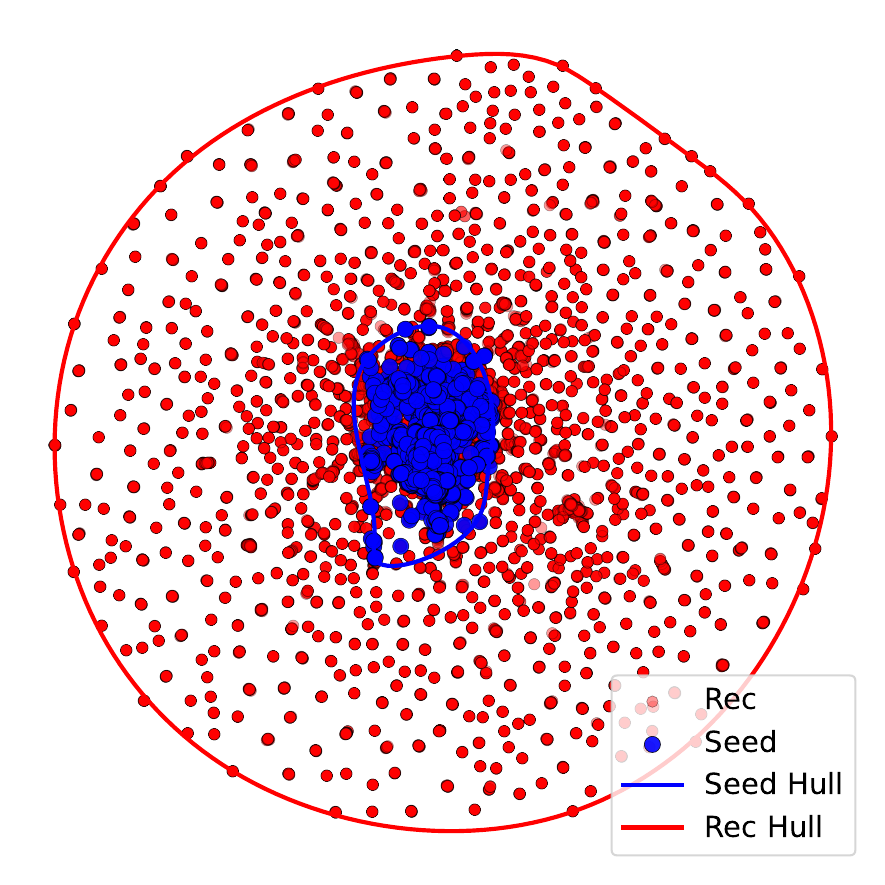}
        \caption{Visual embeddings of keyframes (SCS)}
        \label{umap-figure-scs}
    \end{subfigure}
    \hfill
    \begin{subfigure}[b]{0.48\columnwidth}
        \centering
        \includegraphics[width=\linewidth]{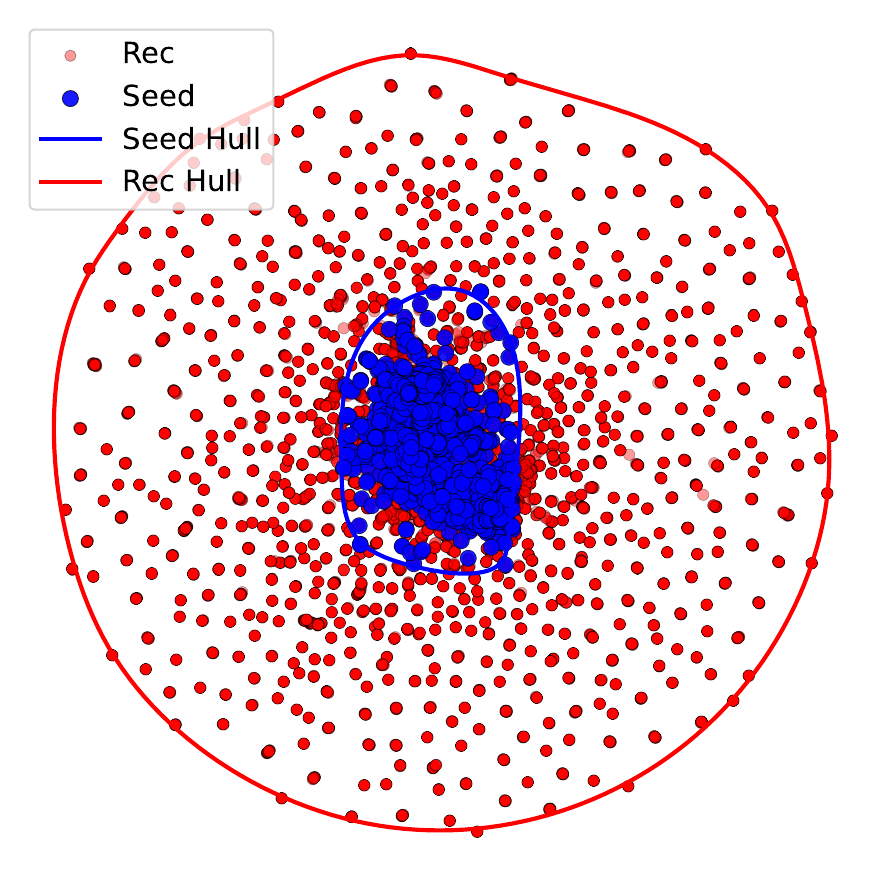}
        \caption{Caption embeddings of keyframes (SCS)}
        \label{umap-caption-figure-scs}
    \end{subfigure}
    
    \vspace{1em}
    
    \begin{subfigure}[b]{0.48\columnwidth}
        \centering
        \includegraphics[width=\linewidth]{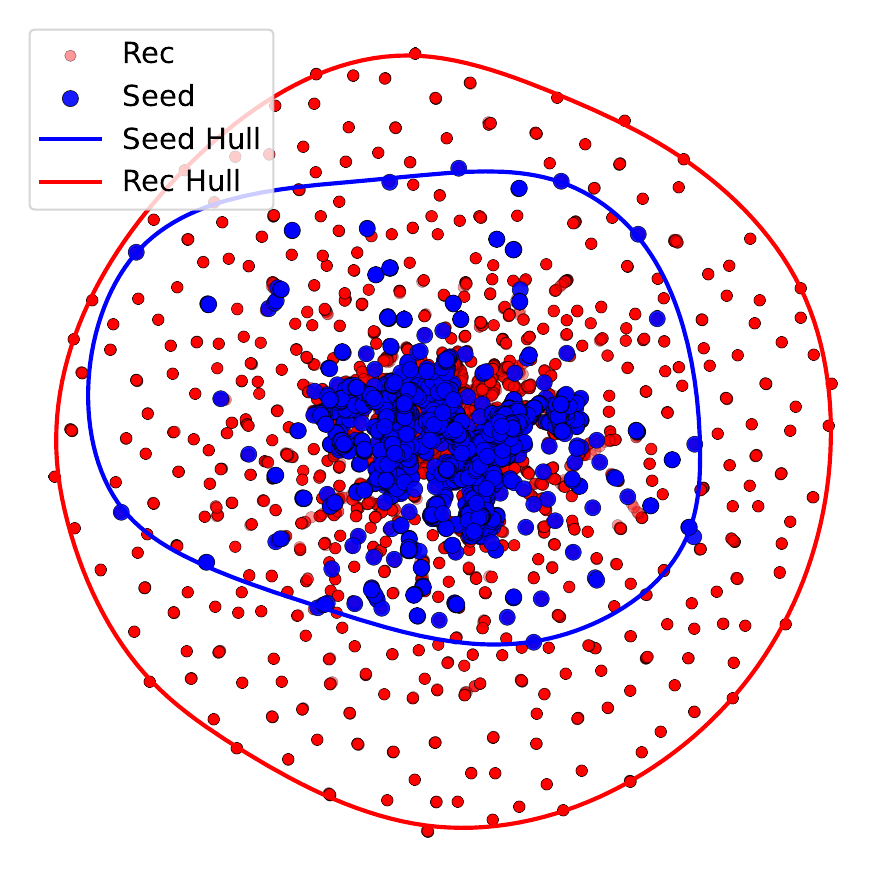}
        \caption{Visual embeddings of keyframes (General)}
        \label{umap-figure-general}
    \end{subfigure}
    \hfill
    \begin{subfigure}[b]{0.48\columnwidth}
        \centering
        \includegraphics[width=\linewidth]{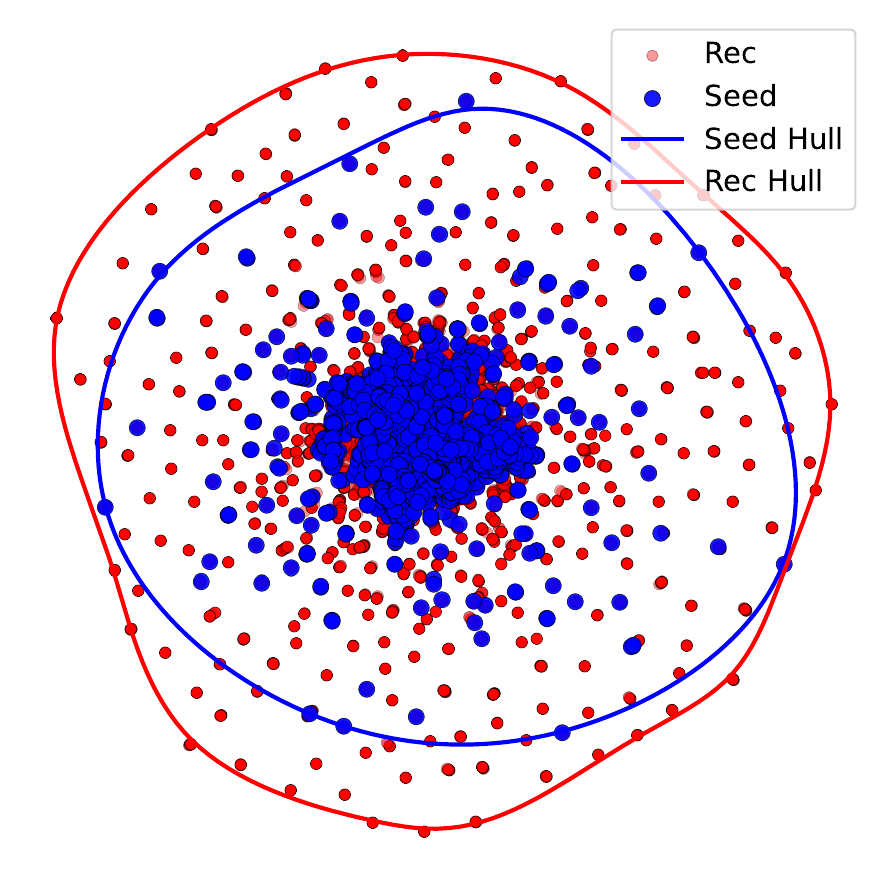}
        \caption{Caption embeddings of keyframes (General)}
        \label{umap-caption-figure-general}
    \end{subfigure}
    
    \caption{UMAP projections of keyframe and caption embeddings for South China Sea and general YouTube content. Each pair compares the distribution between seed and recommended videos.}
    \label{fig:umap_all}
\end{figure}

Figure~\ref{fig:umap_all} shows UMAP projections for the politically sensitive \textit{South China Sea} (SCS) dataset and general YouTube content. In the SCS domain (Figures~\ref{umap-figure-scs} and~\ref{umap-caption-figure-scs}), seed videos (blue points and hulls) form a tight cluster, indicating a focused topical anchor. In contrast, recommended videos (red points) are more dispersed and frequently fall outside the seed region, suggesting that YouTube’s algorithm introduces content that diverges in theme, style, or presentation. This supports \textbf{RQ1}, confirming that recommendations shift away from their seed content, particularly in politically sensitive contexts.

For general YouTube content (Figures~\ref{umap-figure-general} and~\ref{umap-caption-figure-general}), the seed distribution is broader, yet recommendations remain closer and more overlapping, indicating a more stable and consistent recommendation pattern. This aligns with \textbf{RQ3}, showing that the degree of shift varies by domain: sensitive topics experience greater divergence, while general content remains more coherent.

Keyframe and caption embeddings exhibit parallel patterns, with independently derived clusters reinforcing one another. This supports \textbf{RQ2}, demonstrating that keyframes and their captions together provide interpretable signals of recommendation behavior and potential bias.

To support the visual analysis, we compute quantitative metrics that capture structural differences between seed and recommended video embeddings. Table~\ref{tab:cluster_stats_multicol} shows that in the general dataset, recommendation variance increases moderately (e.g., 46.91 for captions and 59.69 for frames). In contrast, the SCS dataset exhibits substantially higher spread, with variances of 79.13 (captions) and 89.13 (frames), as well as greater intra-cluster distances (15.63 and 16.74), indicating more dispersed and thematically inconsistent recommendations. These differences suggest that recommendations originating from politically sensitive content tend to drift more broadly in both visual and textual dimensions.

Table~\ref{tab:divergence_tests} quantifies the divergence between the South China Sea and general YouTube datasets. Using Jensen-Shannon Divergence and Wasserstein Distance to capture global distributional shifts, and $\Delta$ metrics to assess structural spread, we observe consistently higher scores in frame space (e.g., JSD = 0.889, $\Delta$ variance = 0.909). This suggests that visual features diverge more sharply than textual ones across the two domains. While caption embeddings also show notable divergence, the differences are less pronounced than in the visual space. One interpretation is that visual framing, such as imagery, style, or scene composition, varies more between sensitive and general content than caption-level semantics. This further underscores the value of analyzing both modalities and highlights how visual presentation may subtly reflect topic sensitivity in algorithmic recommendations.

\begin{table}[ht]
\centering
\caption{Normalized divergence scores comparing South China Sea and general YouTube datasets across embedding spaces. Frame embeddings show stronger divergence than captions.}
\label{tab:cluster_stats_multicol}
\resizebox{\columnwidth}{!}{%
\begin{tabular}{lcccc}
\toprule
\textbf{Metric} & \multicolumn{2}{c}{\textbf{General}} & \multicolumn{2}{c}{\textbf{South China Sea (SCS)}} \\
\cmidrule(lr){2-3} \cmidrule(lr){4-5}
 & \textbf{Caption} & \textbf{Frame} & \textbf{Caption} & \textbf{Frame} \\
\midrule
Seed Variance           & 7.82  & 12.86 & 4.29  & 4.23  \\
Rec Variance            & 46.91 & 59.69 & 79.13 & 89.13 \\
Seed Intra Dist         & 4.70  & 6.11  & 3.50  & 3.48  \\
Rec Intra Dist          & 11.55 & 13.32 & 15.63 & 16.74 \\
\bottomrule
\end{tabular}%
}
\end{table}

\begin{table}[ht]
\centering
\caption{Normalized divergence scores for SCS and General datasets. Lower values indicate minimal difference; higher values reflect stronger divergence between seed and recommendation distributions. JSD and Wasserstein capture global shifts, while $\Delta$ metrics reflect clustering changes.}

\label{tab:divergence_tests}
\begin{tabular}{lcc}
\toprule
\textbf{Metric} & \textbf{Caption Space} & \textbf{Frame Space} \\
\midrule
Jensen-Shannon Divergence       & 0.708 & 0.889 \\
Wasserstein Distance (scaled)   & 0.552 & 0.602 \\
Normalized $\Delta$ Variance    & 0.662 & 0.909 \\
Normalized $\Delta$ Intra-Dist  & 0.629 & 0.870 \\
\bottomrule
\end{tabular}
\end{table}

Together, these visual and quantitative results offer a consistent, multimodal perspective on how recommendation behavior varies across domains. Our keyframe-based approach captures these shifts effectively and presents a scalable, interpretable method for auditing content dynamics in short-form video platforms.

\section{Conclusion}

In this study, we introduced a keyframe-based framework for analyzing YouTube Shorts recommendation behavior using both visual and textual signals. By extracting perceptually salient frames and generating semantic embeddings, we traced how content shifts through the recommendation pipeline. Our findings reveal substantial content drift in politically sensitive topics such as the South China Sea, in contrast to the more stable recommendation patterns observed in general content. The key contribution of this work is demonstrating that keyframes serve as a scalable and interpretable proxy for detecting algorithmic bias without requiring full video processing. While this study focuses on two domains, future research could broaden the scope to include additional topics and incorporate complementary analyses such as topic modeling, sentiment tracking, or engagement signals. Overall, our results underscore how recommender systems can subtly shape content exposure and reinforce the importance of transparency in algorithmic media distribution.


\section*{GenAI Usage Disclosure}

The authors used generative AI tools as part of the research methodology. Specifically, the LLaMA-3.2-11B-Vision-Instruct model was used to generate textual captions from video keyframes for content analysis. No generative AI tools were used in the writing, coding, or data collection processes beyond this purpose.

\begin{acks}

This research is funded in part by the U.S. National Science Foundation (OIA-1946391, OIA-1920920), U.S. Office of the Under Secretary of Defense for Research and Engineering (FA9550-22-1-0332), U.S. Army Research Office (W911NF-23-1-0011, W911NF-24-1-0078, W911NF-25-1-0147), U.S. Office of Naval Research (N00014-21-1-2121, N00014-21-1-2765, N00014-22-1-2318), U.S. Air Force Research Laboratory, DARPA, the Australian DoD Strategic Policy Grants Program, Arkansas Research Alliance, the Jerry L. Maulden/Entergy Endowment, and the Donaghey Foundation at UA Little Rock. Opinions are the authors’ own and do not necessarily reflect the funders; we gratefully acknowledge their support.
\end{acks}

\printbibliography




\end{document}